\documentstyle[prl,aps,epsfig,floats]{revtex}

\begin{document}
\draft
\twocolumn[\hsize\textwidth\columnwidth\hsize\csname
@twocolumnfalse\endcsname

%\begin{frontmatter}
\title{Possible Hints and Search for Glueball Production in Charmless Rare $B$ 
Decays}

\author{Chun-Khiang Chua, Wei-Shu Hou and Shang-Yuu Tsai\\}
\address{Department of Physics, National Taiwan University,
Taipei, Taiwan 10764, R.O.C.}
\date{\today}
\maketitle

\begin{abstract}

Recent data on $B\to p\bar pK$, $K^0 \pi\pi$ and $KK\bar K$
hint at a $\sim 2.3$ GeV object recoiling against a kaon.
This could be the narrow state observed in $J/\psi \to \gamma\xi$.
Nonobservation in $p\bar p$ annihilation implies
${\cal B}(\xi \to p\bar p) \sim$ few $\times 10^{-3}$,
consistent with $\eta_c$ and $J/\psi$ decays, 
but there are actual hints in $p\bar p \to \phi\phi$ 
and $pp\to p\pi^+\pi^-\pi^+\pi^-p$.
Simple modeling shows 
${\cal B}(B \to \xi K){\cal B}(\xi \to p\bar p) \sim 1\times 10^{-6}$,
appearing as a spike in the $p\bar p$ spectrum,
with $\sim$ 30 events per 100 fb$^{-1}$;
modes such as $KK_sK_s$, $K\phi\phi$, $K4\pi$ ($Kf_2\pi\pi$) 
etc. should be explored.
The underlying dynamics of $g^*\to g\xi$ is analogous to
$g^*\to g\eta^\prime$ or gluon fragmentation.
Discovery of sizable $B\to \xi K$ could be useful for CP violation studies.

\end{abstract}

]
%\begin{keyword}
%\PACS 13.25.Hw\sep 13.40.Gp\sep 14.20.Dh
%\end{keyword}
%\end{frontmatter}
%
%\preprint{\vbox{\hbox{}}}
%
%\preprint{\vbox{\hbox{}}}
%
%\preprint{\vbox{\hbox{}}}
%

%%%%%%%%%%%%%%%%%%%%%%%%%%%%%%%%%%%%%%%%%%%%%%%%%%%%%%%%%%%%%%%%%%

%\section{Introduction}\label{Intro}

The existence of glueballs as bound states of gluons, 
the gauge bosons of QCD, has been conjectured ever since the advent of 
QCD as the fundamental theory of the strong interaction. 
Alas, it is a unique feature of {\it nonabelian} gauge theories 
that has yet to be unequivocally tested. 
The main obstacle to identifying glueballs is their possible 
$q\bar q$ admixture, which allows the candidates 
to hide in the richness of $q\bar q$ resonances.
Advances in lattice gauge theories suggest the lowest lying glueballs to be 
the $0^{++}$ scalar  with $m_G \sim$ 1.4--1.8 GeV
and $2^{++}$ tensor with $m_\xi \sim$ 1.9--2.3~GeV,
%lying in the mass range of 1.4--1.8 GeV and 1.9--2.3~GeV,
while the $0^{-+}$ glueball $P$ is another 150 MeV heavier \cite{Morning}.
%We denote these glueballs as $G$, $\xi$ and $P$, respectively.

Radiative $J/\psi\to \gamma gg \to \gamma+$hadrons decay
is a prime hunting ground for glueballs.
The narrow state $\xi$ with width 23 MeV, 
called $f_J(2220)$ by the Particle Data Group (PDG) \cite{PDG},
was discovered \cite{mrk3} by the MARK~III experiment in such decays.
The BES collaboration confirmed \cite{Bai} the $\xi$ signal in
$J/\psi \to \gamma\pi^+\pi^-$, $\gamma K^+K^-$, 
$\gamma K_S^0K_S^0$, $\gamma p\bar p$ at
$(5.6\pm 2.7)\times 10^{-5}$, $(3.3\pm 2.0) \times 10^{-5}$, 
$(2.7\pm 1.4)\times 10^{-5}$, $(1.5\pm 0.8)\times 10^{-5}$, respectively, 
as well as $J/\psi \to \gamma\pi^0\pi^0 \sim (4.5\pm 2.9)\times 10^{-5}$,
where errors have been combined conservatively.
Null results in $\gamma\gamma \to \xi$ search \cite{twophoton}
strengthen the glueball interpretation.

The $\xi\to p\bar p$ mode stimulated scans of 
$p\bar p$ annihilation around 2230 MeV,
resulting in the limits of
$p\bar p\to K_S^0K_S^0$, $\phi\phi$, $\pi^0\pi^0$, $\eta\eta
< 7.5\times 10^{-5}$, $6\times 10^{-5}$~\cite{Evangelista},
 $6\times 10^{-5}$, $4\times 10^{-5}$~\cite{Amsler}, 
respectively.
Combining with the BES result, one finds \cite{Amsler}
that ${\cal B}(\xi \to p\bar p) \lesssim 5\times~10^{-3}$,
and
\begin{eqnarray}
{\cal B}(J/\psi \to \gamma\xi) & \gtrsim & 2.9\times~10^{-3},
 \label{gammaxi}
% \\ \label{xipp}
\end{eqnarray}
which seems to support the glueball interpretation.
However, the nonobservation in quite a few $p\bar p$ annihilation modes
has lead to doubt \cite{Close01} of the very existence of $\xi$.

With this impasse, it is desirable to open up 
new avenues for exploration.
The charmless $b\to sg^*$ process could be \cite{AS2,HHH}
viable ground for glueball search.
This was stimulated in part by the CLEO observation of large
$B\to \eta^\prime K \sim 8\times 10^{-5}$ \cite{JimS} and
$\eta^\prime + X_s \gtrsim 6\times 10^{-4}$ \cite{Browder},
which were interpretted \cite{AS,HT} as related to the large glue content
of $\eta^\prime$ via the gluon anomaly.
The $b\to sg^*$ transition, followed by the anomaly inspired 
effective $g^*\to g\eta^\prime$ coupling, could account for \cite{Browder,HT}
the semi-inclusive $m_{X_s}$ spectrum.
Replacing $\eta^\prime$ by a glueball may be even 
more effective \cite{AS2,HHH}.
%The exclusive $B\to \eta^\prime K$ modes have recently been confirmed
%by B Factory experiments \cite{Paoti}, 
%while the inclusive studies are picking up \cite{etapX}.
In this Letter we point out possible hints for 
$B\to \xi K$ decay in the $B\to p\bar pK$, $K_S \pi^+\pi^-$ 
and $K^+ K^+K^-$ modes newly observed by Belle, 
and discuss directions for further study.

%%%%%%%%%%%%%%%%%%%%%%%%%%%%%%%%%%%%%%%%%%%%%%%%%%%%%%%%%%%%%%%%%%%%%%%
\begin{figure}[b!]
\begin{center}
\epsfig{figure=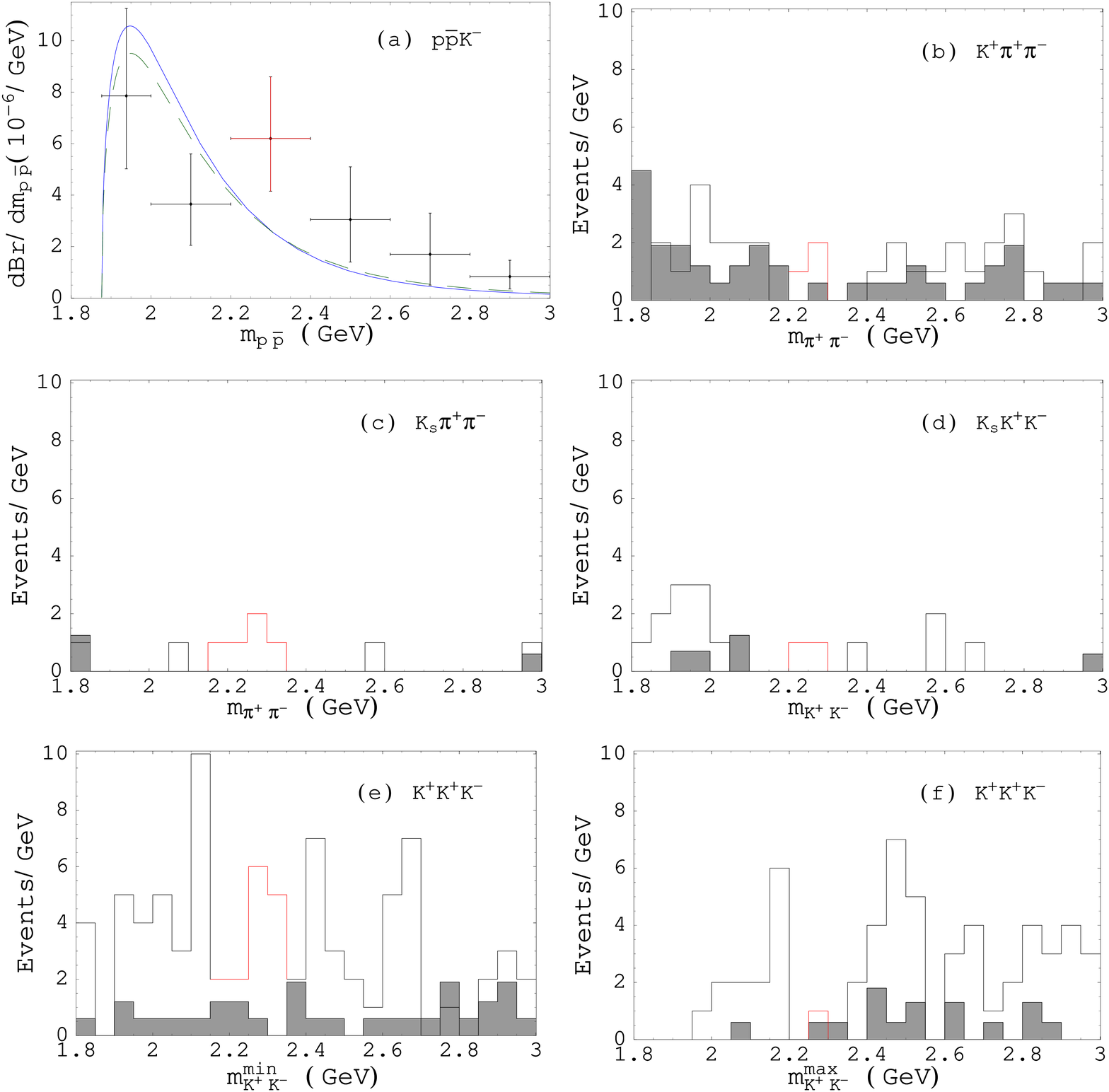,width=3.5in}
\end{center}
\caption{Spectra for $B \to$ (a) $p\bar pK$; 
(b) $K\pi\pi$ ($m_{K\pi} > 2$ GeV);
(c) $K_S\pi\pi$; (d) $K_SK\bar K$; and 
(e), (f) for $KK\bar K$ vs. $m_{K\bar K}^{\rm min}$, 
$m_{K\bar K}^{\rm max}$ ($m_{K\bar K}^{\rm min} > 1.1$ GeV), 
respectively.}
\label{spectra}
\end{figure}
%%%%%%%%%%%%%%%%%%%%%%%%%%%%%%%%%%%%%%%%%%%%%%%%%%%%%%%%%%%%%%%%%%%%%%%

Let us first present the case for charmless $B$ decays.
The $B\to p\bar pK$ decay \cite{ppK} is the 
first ever charmless {\it baryonic} mode to be observed. 
While modeling the $m_{p\bar p}$ spectrum
by a QCD motivated threshold enhancement,
we noted a hint for a $\sim 2.3$ GeV peak.
The data (fitted ${\cal B}$) and our modeling \cite{CHT} are 
plotted in Fig.~1(a).
Threshold enhancement is apparent, in line with our prediction \cite{rhopn}
for $B\to \rho p\bar n$ before the discovery of $B\to p\bar pK$.
However, some excess $\sim$ 7--10 events is noticeable in the third, 
i.e. 2.2--2.4 GeV bin \cite{Minzu}, 
amounting to $\sim 0.6$--$1\times 10^{-6}$ in rate,
which we could not accommodated in our simple threshold model.
Motivated by this, we find evidence in 
a few (but not all) other 3-body channels as well.

The $B\to K^+\pi^+\pi^-$ mode
observed by Belle \cite{Khh} is plotted in Fig. 1(b), 
with a cut of $m_{K^+\pi^-} > 2$ GeV to suppress background.
Despite some activity above 2~GeV,
there is not much excess at 2.2--2.3~GeV.

The $B\to K_S\pi^+\pi^-$, $K_SK^+K^-$ modes, 
also observed by Belle \cite{HC},
are plotted in Figs. 1(c) and (d), respectively.
The spectrum for $m_{\pi^+\pi^-} >$ 2 GeV is very clean, with 
a striking cluster at 2.3 GeV, albeit with only 5 events.
The $m_{K^+K^-}$ spectrum has $\sim$ 2 events in the same region
(but a prominent cluster at $\sim 1.95$ GeV).
In all, ${h^+h^-}$ has about 7 events, and folding in efficiencies, 
we find the average over $K_S\pi^+\pi^-$, $K_SK^+K^-$ rates in 
the cluster region is $\sim 2.5\times 10^{-6}$.
The comparison with $p\bar pK$ case is consistent with
the BES observation.

Turning to $B\to K^+K^+K^-$ \cite{Khh}, 
we plot $m_{K^+K^-}^{\rm min}$ and $m_{K^+K^-}^{\rm max}$
 (for $m_{K^+K^-}^{\rm min} > 1.1$ GeV) spectra in Figs. 1(e) and (f).
The $m_{K^+K^-}^{\rm min}$ spectrum above 2 GeV is 
quite sizable and rich with structure, 
like Fig. 1(b) amplified but with much less background.
This decay is expected to arise solely from the $b\to ss\bar s$ penguin.
One has $\sim$ 10 events each 
at 2.3, 2.45 and 2.65 GeV, 
and $\sim$ 20 events at 1.9--2.15 GeV,
the latter similar to $K_SK^+K^-$.
For $m_{K^+K^-}^{\rm max}$ one has $\sim$ 11, 14, 6 events
respectively at 2.1, 2.45 and 2.65 GeV,
but no 2.3 GeV cluster.
Folding in efficiencies, we find a rate of 1.7 to 3.4 $\times 10^{-6}$,
again consistent with BES and with $K_Sh^+h^-$.
We caution, however, that identical particle effects,
reflected in two possible $K^+K^-$ pairings, smear the plots.

To summarize,
there is some evidence for a 2.2--2.3 GeV ``state" recoiling 
against a kaon in $p\bar pK$, $K_Sh^+h^-$ and $K^+K^+K^-$ channels,
which could be the $\xi$ glueball candidate.
The $\sim$ 2.45 or 2.65 GeV objects might be the pseudoscalar $P$
(or a scalar excitation \cite{Morning});
there is also some excess in these regions for $p\bar pK$ %\cite{Ptopp}
 (Fig. 1(a)).
The absence in $K^+\pi^+\pi^-$ is worrisome,
but, besides larger background (hence extra cut),
there are also amplitude level complications, 
such as a slower fall-off in $m_{\pi\pi}$ vs. $m_{p\bar p}$, 
the tree contribution (in contrast to $K_Sh^+h^-$), 
and multiple interfering resonances.
We conclude that glueballs may emerge in 
higher statistics studies of charmless rare $B$ decays,
and wish to survey what we know about, and how to
gain access to, such glueballs.

The $J/\psi \to \gamma K^+K^-$, $\gamma K_S^0K_S^0$ 
numbers from BES \cite{Bai} are 
slightly below MARK III results\cite{mrk3}, 
while the $p\bar p$ number is just below the bound of $2\times 10^{-5}$.
But the $\pi^+\pi^-$ number is $\sim$ factor 3 {\it above}
the MARK III bound of $2\times 10^{-5}$.
Since there are two structures adjacent to the $\pi^+\pi^-$ peak
in BES data, the actual rate is probably smaller.
If the 2.2--2.3 GeV ``signal" in 
$B\to p\bar pK$, $K_Sh^+h^-$ and $K^+K^+K^-$ is due to the $\xi$,
our discussion above indicates that 
$J/\psi \to \gamma\pi^+\pi^-$, $\gamma K^+K^- \sim (3$--$4)\times 10^{-5}$
would be more consistent, 
hence $J/\psi \to \gamma K_S^0K_S^0\sim (1.5$--2)$\times 10^{-5}$, 
slightly lower than BES.
The BES result for 
$J/\psi \to \gamma\pi^0\pi^0$ \cite{Bai}
is almost twice larger than implied by their 
$J/\psi \to \gamma\pi^+\pi^-$, and was not used
in the PDG estimate \cite{PDG} of $J/\psi \to \gamma\pi\pi$.

An intriguing recent result has come from CLEO.
Based on 61.3 pb$^{-1}$ data $\cong$ 1.45 million $\Upsilon(1S)$ mesons, 
CLEO reports \cite{Upsilon} 1, 1, 2 events 
within $\pm 34$ MeV of 2234 MeV in 
$\Upsilon \to \gamma\pi^+\pi^-$, $\gamma K^+K^-$, $\gamma p\bar p$,
respectively, with background expected at 0.12, 0.21, 0.28; 
a {\it lower bound} of 
$\Upsilon \to \gamma\xi \to \gamma p\bar p > 0.5\times 10^{-6}$ is obtained.
CLEO chose to drop this by allowing for larger background.
However, scaling \cite{Upsilon} 
the BES $J/\psi \to \gamma p\bar p$ result by
$(Q_b^2m_c^2\Gamma_{J/\psi}/Q_c^2m_b^2\Gamma_{\Upsilon}) \sim 0.04$ gives 
$\Upsilon \to \gamma p\bar p = (0.6\pm 0.3)\times 10^{-6}$,
right in the ballpark.
We mention that CLEO has just finished \cite{JimA} taking
1.3 fb$^{-1}$ data on the $\Upsilon(1S)$, i.e. a 21-fold increase,
and we may see the $\xi$ popping up in radiative $\Upsilon$ decays,
with 10 to 40 events in the $\pi^+\pi^-$, $K^+K^-$ and $p\bar p$ 
(and other) modes in the near future.

It is the $p\bar p$ annihilation experiments which 
cast doubt on the existence of $\xi$.
These experiments were stimulated by the BES observation of $\xi \to p\bar p$
to scan around 2230 MeV,
before CERN Lower Energy Antiproton Ring (LEAR) shutdown in 1996. 
The results were all negative.
The conservative conclusion is that
$\xi \to \pi^+\pi^-$, $K^+K^-$, $K_S^0K_S^0$, $p\bar p$, 
$\phi\phi$, $\pi^0\pi^0$, $\eta\eta$ are all $\lesssim 1\%$.
But, together with the {\it narrow} $\Gamma_\xi \sim 20$ MeV, 
the stated doubt \cite{Close01} grew with time.
We offer a critique of the situation.

First,
two body decays of $\xi\lesssim 1\%$ is not surprising.
The $\eta_c$ and $J/\psi$ decays via $gg$ and $ggg$, and their
$p\bar p$ rates are 0.12\% and 0.21\% \cite{PDG}, respectively.
If the $\xi$ is the $2^{++}$ two-gluon glueball, having 
${\cal B}(\xi\to p\bar p) \sim $ few $\times 10^{-3}$ %(cf. Eq. (2)) 
seems just right.
Second, 
a 20 MeV width for a lowest lying 2.2--2.3 GeV two-gluon glueball
is also not unreasonable.
On one hand, the ``$\sqrt{\rm OZI}$" rule \cite{Hou:1996qk}, 
i.e. taking the geometric mean of 
the few MeV width of $\eta_c$ (scaled down to 2 GeV)
and the few hundred MeV width of a typical 2 GeV meson 
gives 10--50 MeV.
On the other hand, the near ideal mixing of
$f_2(1270)$--$f_2^\prime(1525)$ system implies \cite{HK} that
the relevant lowest lying glueball, the $\xi$, 
would be relatively free of $q\bar q$ content,
hence the above narrowness argument holds.
Third,
the lower bound of Eq. (1) is not unreasonable if $\xi$ is really a glueball, 
but the large ${\cal B}(J/\psi \to \gamma\xi)$ is a bit overstated.
It arises from combining 
the BES result on $J/\psi \to \pi^0\pi^0$ \cite{Bai}
with the nonobservation of $p\bar p \to \pi^0\pi^0$ \cite{Amsler}.
As we noted, the BES result for $\pi^0\pi^0$
is likely a factor of 2 to 3 too large.
 
With these points, 
it should be clear that $\xi$ is still viable.
We now argue that there is in fact some evidence coming from
$p\bar p$ annihilation or $pp$ collisions.

Although the JETSET experiment did not observe 
a narrow $\xi$ in $p\bar p \to \phi\phi$ channel, 
they did find \cite{Evangelista} a broad structure just above threshold.
In fact, further partial wave analysis \cite{Palano} 
found $2^{++}$ dominance, and a resonance behavior in $2^+_{D0}$
($D$-wave with $\phi\phi$ spin zero):
a Breit-Wigner structure with phase motion vs. $2^+_{D2}$, 
consistent with $m \cong 2231$ MeV and $\Gamma \cong 70$ MeV.
From Fig.~6 of Ref.~\cite{Palano}, 
comparing $2^+_{D0}$ with $2^+_{D2}$, $2^+_{S2}$ waves, 
we note that it may be better to fit with {\it two} Breit-Wigner resonances
(or one resonance with a broad underlying structure).
We believe the JETSET data does not preclude 
a narrow resonance at 2.2 GeV.

There is another hint in central hadron production.
The empirical ``$dP_T$" glueball filter \cite{Close} is
defined as the difference between the transverse momenta of 
e.g. the outgoing protons in $pp \to p X p$;
$dP_T \to 0$ enhances glueball probability of $X$.
Using data from WA102 experiment with $X = \pi^+\pi^-\pi^+\pi^-$, 
it was shown that the $f_1(1285)$ prominent for larger $dP_T$ 
all but disappeared for $dP_T < 0.2$ GeV, 
while the glueball candidate $f_0(1500)$ is retained.
From Fig. 3(c) of Ref. \cite{Close}, however, we find 
a remarkable single-bin (2320--2340 MeV) spike, absent for $dP_T > 0.2$ GeV, 
but popping up for $dP_T < 0.2$ GeV. 
{\it With $\simeq 100$ events on $\simeq 360$,
it constitutes a $>5\sigma$ fluctuation}.
The detector resolution is $\sim 12$ MeV \cite{WA102}
hence the spike seems genuine.
A broader structure exists at 2430 MeV.
Subsequent spin analysis (Fig. 3(f) of second paper of Ref. \cite{WA102}) 
also show a ``spike" at 2240--2280 MeV,
and a broader structure at 2400 MeV,
all in the $2^{++}$ channel of $f_2\pi\pi$.
By analogy with the 
large $\eta_c\to \eta^{(\prime)}\pi\pi \sim (4$--5)\% \cite{PDG},
$\xi \to f_2\pi\pi$ could be a major decay mode.
These features should be investigated further.

%%%%%%%%%%%%%%%%%%%%%%%%%%%%%%%%%%%%%%%%%%%%%%%%%%%%%%%%%%%%%
\begin{table}[t!]
\caption{
Branching ratios ($\times 10^{-6}$) of the
$\bar B\to \xi K^-$, $p\bar pK^-$, $p\bar p\bar K^0$ 
and $p\bar p \pi^-$ modes with $\Gamma_\xi=23$~MeV.
 %and take $\phi_3=90^\circ$. 
The first two numbers for ${\mathcal B}(p\bar pK^-)$ correspond to 
the upper and the lower curves of Fig.~\ref{spectra}~(a) [16], 
respectively.
}
\begin{center}
\begin{tabular}{cccc}
$f_{B\xi K}$ & $0$ & $0.014~(0.015)$ & $-0.014~(-0.016)$ \\
${\mathcal B}(B^-\to \xi K^-)$ & $0$ & $220~(260)$ & $240~(300)$ \\
\hline
${\mathcal B}(B^-\to p\bar pK^-)$ & $3.4~(3.3)$ & \multicolumn{2}{c}{$4.3$}\\
\hline
${\mathcal B}(\bar B^0\to p\bar p\bar K^0)$ & $3.3~(0.5)$ & $4.1~(1.4)$ &
$4.1~(1.5)$ \\
${\mathcal B}(B^-\to p\bar p\pi^-)$ & $2.1~(2.1)$ & $2.1~(2.1)$ &
$2.1~(2.1)$ \\
\end{tabular}
\end{center}
\end{table}
%%%%%%%%%%%%%%%%%%%%%%%%%%%%%%%%%%%%%%%%%%%%%%%%%%%%%%%%%%%%%

%%%%%%%%%%%%%%%%%%%%%%%%%%%%%%%%%%%%%%%%%%%%%%%%%%%%%%%%%%%%%%%%%%%
\begin{figure}[t!]
\vspace{-1cm}
\begin{center}
\epsfig{figure=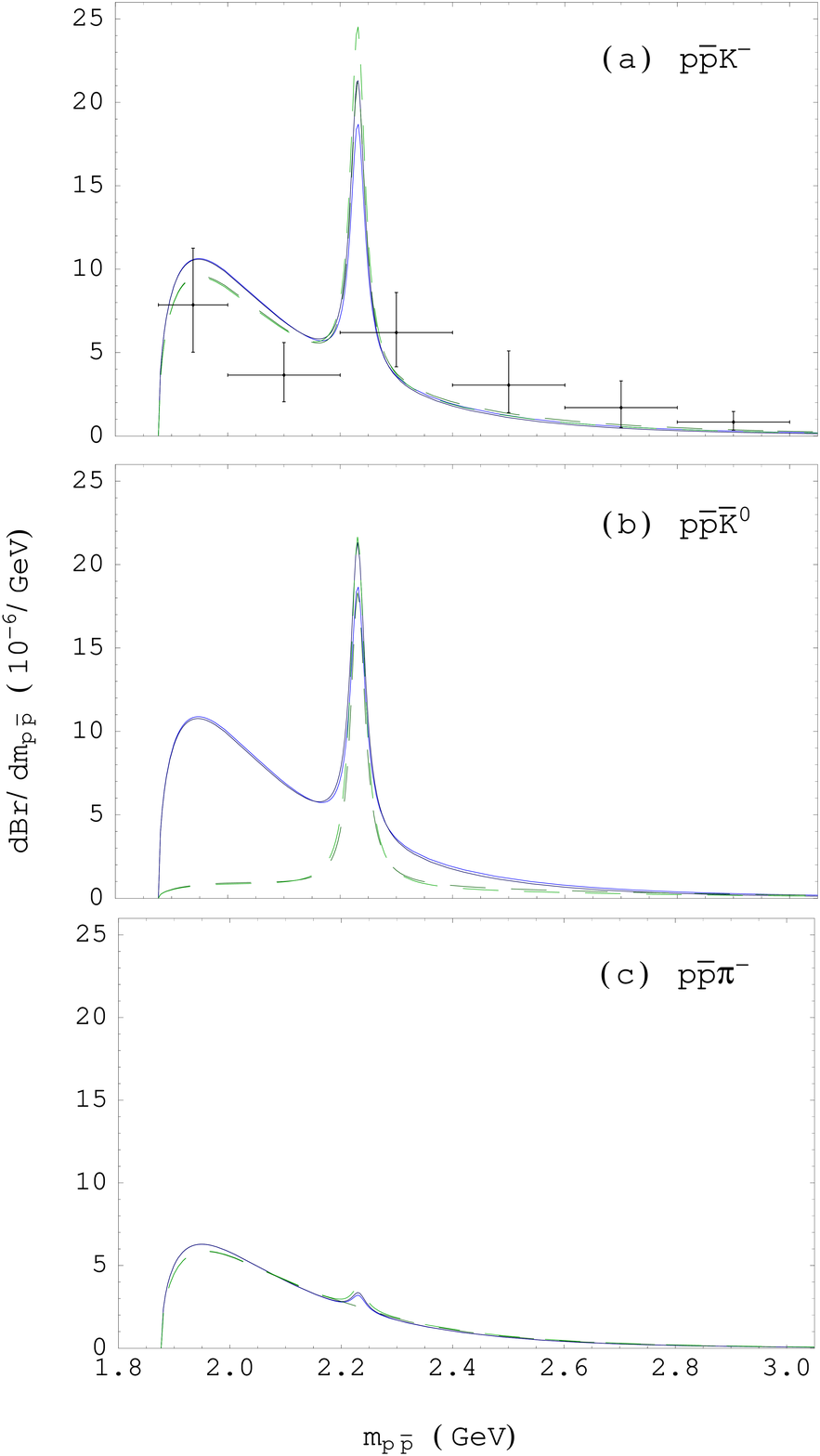,width=3in,height=4.5in}
\end{center}
\caption{Modeling of $\bar B\to p\bar pK^-$, $p\bar p \bar K {}^0$ and
$p\bar p\pi^-$ spectra with $\xi$ spike. 
For illustration we plot the $\Gamma_\xi=70$~MeV case.
The upper (lower) curves correspond to the upper (lower)
one in Fig. 1 (a) (from Ref. [16]).
}
\label{xiK}
\end{figure}
%%%%%%%%%%%%%%%%%%%%%%%%%%%%%%%%%%%%%%%%%%%%%%%%%%%%%%%%%%%%%%%%%%%

We now turn to simple modeling of the $B\to p\bar pK$ ``bump"
assuming a $2^{++}$ glueball state.
That is, we have a $B\to \xi K(\pi)$ transition
governed by
\begin{equation}
{G_F \over \sqrt{2}} V_{tb}V_{ts(d)}^* \, f_{B\xi K(\pi)} \,
 \varepsilon^*_{\mu\nu} \, p_B^\mu p_{K(\pi)}^\nu,
\end{equation}
where we factor out the quark mixing factor appropriate for the
underlying $b\to s(d)$ penguin,
The $\xi \to p\bar p$ transition is governed by
%\begin{equation}
$-g_1^{\xi p\bar p} \, \varepsilon_{\mu\nu}
 \bar u \gamma^\mu p_{\bar p}^\nu v$,
%\end{equation}
where a less effective $p_{p}^\mu p_{\bar p}^\nu$ term is dropped.
For given $\Gamma_\xi$, $g_1^{\xi p\bar p}$ is fixed by
${\cal B}(\xi \to p\bar p)\sim 5\times~10^{-3}$. 
Together with the fits in Fig. 1(a), $f_{B\xi K}$ is fixed
(its sign determines interference; we ignore relative strong phase) 
to reproduce ${\mathcal B}(B\to p\bar pK)=4.3\times10^{-6}$.
%We obtain $g_1^{\xi p\bar p}=0.15$, up to a simple scaling if other value of 
%${\cal B}(\xi \to p\bar p)$ is used. 
The results for the 
$\bar B\to \xi K^-,\,p\bar pK^-,\,p\bar p\bar K^0$ and 
$p\bar p \pi^-$ modes are given in Table I for $\Gamma_\xi = 23$ MeV,
and their spectra in Fig.~2 for
$\Gamma_\xi =70$ MeV for sake of illustration.
The $p\bar p\bar K^0$ case depends on the threshold dynamics, 
while the $\xi$ is far less prominent in $p\bar p \pi^-$ 
because it is tree dominant.

The underlying dynamics of $B\to \xi K$ is rather 
analogous to that proposed for $B\to \eta^\prime K$ and $\eta^\prime X_s$.
We have factored out in Eq. (2) $G_F/\sqrt{2}$ and
the $V_{tb}V_{ts}^*$ quark mixing factor coming from the penguin loop, 
as illustrated in Fig. 3.
The $g^*\to g\xi$ coupling is the heavy blob,
similar to $g^*\to g\eta^\prime$ via the gluon anomaly \cite{AS,HT}, 
which gives rise to the glue-content of $\eta^\prime$.
As argued in Ref. \cite{HHH} (see also \cite{AS2}) for the case of $P$,
the production of a {\it bona fide} glueball in a similar fashion 
may be even more effective.
How the $sg\bar q$ system evolves into a kaon is not of concern
here since $B \to \eta^\prime K$ is observed \cite{Browder},
and its large strength, recently confirmed by both
Belle and BaBar \cite{Paoti}, is still not explained by theory.
Thus, it is plausible that $B\to \xi K > B \to \eta^\prime K$ 
and could be $\gtrsim 10^{-4}$.
There has been some perturbative arguments for $1/q^2$ damping
of the effective $g^*g\xi$ vertex \cite{Ali},
but since nonperturbative effects 
--- which generate $m_\xi^2 \gg m_\rho^2$ --- 
are bound to enter, we advocate \cite{HT} to leave the case open.
Note that several authors have suggested \cite{fragment} to search for
glueballs in gluon fragmentation.
The $g^*\to g\xi$ process advocated here can be viewed as
such, {\it but at only a few GeV energy}.
This illustrates further the futility to
discard the $g^*g\xi$ vertex by perturbative arguments.

%%%%%%%%%%%%%%%%%%%%%%%%%%%%%%%%%%%%%%%%%%%%%%%%%%%%%%%%%%%%%%%%%%%%%%%%%%
\begin{figure}[t!]
\centerline{{\epsfxsize3in \epsffile{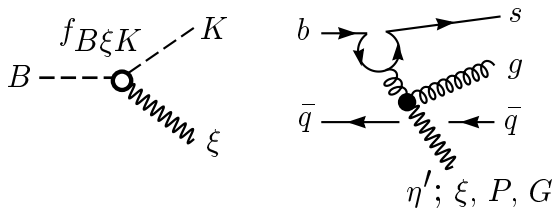}}
	   }
\caption{Illustration for $B\to \xi K$ underlying dynamics. }
 \label{dGdqVrhopi}
\end{figure}
%%%%%%%%%%%%%%%%%%%%%%%%%%%%%%%%%%%%%%%%%%%%%%%%%%%%%%%%%%%%%%%%%%%%%%%

One uniquely interesting feature for studying 
glueball production in charmless $B$ decays is 
the potential it offers for studying $CP$ violation \cite{AS2,HT}.
On one hand, the penguin loop implies sensitivity for 
new physics beyond the Standard Model,
e.g. via the dipole $bsg$ coupling.
On the other hand, if $B\to \xi K$ is really at 
a few $\times 10^{-4}$ and $\xi$ is a narrow state,
one could accumulate a large number of modes
and gain in statistics.
$CP$ asymmetries could be at 10--30\% level even if
new physics contributes only 10\% in amplitude \cite{AS2,HT}.

From our survey,
additional search modes are: $B\to KK_SK_S$,
$K\phi\phi$, $K4\pi$ (e.g. $Kf_2\pi\pi$),
and perhaps $p\bar p K_2^*$, beyond the ones given in Fig. 1.
Semi-inclusive studies,
i.e. $B \to \xi(\to p\bar p,\ \mbox{etc.}) + X_s$, can also be considered.
One can also search for other glueballs such as $P$ and $G$,
e.g. $B\to PK$, $GK$ via $P\to \eta^{(\prime)}\pi\pi$, $K\bar K\pi$.
At the same time, the $\eta^\prime$ study, both exclusive and inclusive,
including $CP$ violation effects, should be pursued further.

In summary, 
we find indication for a narrow state in 
$B\to p\bar pK$, $K_S\pi\pi$ and $K^+K^+K^-$
recoiling against a kaon.
This could be the $2^{++}$ glueball candidate
found in radiative $J/\psi$ decays with 
mass supported by lattice calculations,
and with tantalizing hints in $p\bar p \to \phi\phi$ 
and $pp \to p\pi^+\pi^-\pi^+\pi^-\bar p$.
Glueballs may emerge in the study of charmless rare $B$ decays,
with confirming evidence from $\Upsilon \to \gamma p\bar p$.
Search for $\xi$ (and $P$) in 
$B\to p\bar pK$, $K^+K^+K^-$, $K_Sh^+h^-$, $K^+K_SK_S$,
$K\phi\phi$, $K4\pi$ should be vigorously pursued,
with an eye towards uncovering new physics sources of $CP$ violation.

\acknowledgments

We thank J. Alexander, Y.B. Hsiung, H.C. Huang, A.~Palano 
and M.Z. Wang for discussions,
and partial support from the 
NSC of R.O.C., the MOE CosPA project, 
and the BCP topical program of NCTS.

%\appendix

%

\end{document}